\documentclass{aa}  

\usepackage{graphicx}
\usepackage[varg]{txfonts}
\usepackage{booktabs}
\usepackage{xcolor}
\usepackage{xspace}
\usepackage{natbib,twoopt}
\usepackage{multirow}
\usepackage{longtable}
\usepackage{ltablex}
\usepackage{epsfig}
\usepackage{threeparttable}
\usepackage[normalem]{ulem}
\usepackage[breaklinks=true]{hyperref}
\bibpunct{(}{)}{;}{a}{}{,}             
\makeatletter
  \newcommandtwoopt{\citeads}[3][][]{\href{http://adsabs.harvard.edu/abs/#3}%
    {\def\hyper@linkstart##1##2{}%
     \let\hyper@linkend\@empty\citealp[#1][#2]{#3}}}
  \newcommandtwoopt{\citepads}[3][][]{\href{http://adsabs.harvard.edu/abs/#3}%
    {\def\hyper@linkstart##1##2{}%
     \let\hyper@linkend\@empty\citep[#1][#2]{#3}}}
  \newcommandtwoopt{\citetads}[3][][]{\href{http://adsabs.harvard.edu/abs/#3}%
    {\def\hyper@linkstart##1##2{}%
     \let\hyper@linkend\@empty\citet[#1][#2]{#3}}}
  \newcommandtwoopt{\citeyearads}[3][][]%
    {\href{http://adsabs.harvard.edu/abs/#3}
    {\def\hyper@linkstart##1##2{}%
     \let\hyper@linkend\@empty\citeyear[#1][#2]{#3}}}
\makeatother

\newcommand{\swift}{{\it Swift}\xspace}

\newcommand{\fast}{{\it FAST}\xspace}

\newcommand{\ergs}{erg s$^{-1}$\xspace }

\newcommand\T{\rule{0pt}{1.8ex}}       
\newcommand\B{\rule[-0.8ex]{0pt}{0pt}}

\newcommand{\igr}{IGR~J00291+5934\xspace}
\newcommand{\maxij}{MAXI~J1957+032\xspace}
\newcommand{\dmu}{\,pc\,cm$^{-3}$}

\begin{document}

   \title{A \emph{FAST} search for radio pulsations during the dormant state of the AMSPs \igr and \maxij}

   \author{A. Marino\inst{1,2,3,4,5}\thanks{\url{marino@ice.csic.es}}, E. Parent\inst{1,2,6}\thanks{\url{parent@ice.csic.es}}, F. Coti Zelati\inst{1,2,7}, M. C. Baglio\inst{7}, A. Papitto\inst{8}, A. Sanna\inst{9}, A. Anitra\inst{10}, C. Kazantsev\inst{11},  N. Rea\inst{1,2}, A. Borghese\inst{12}, L. Burderi\inst{5,9}, T. Di Salvo\inst{10}, C. Espinoza\inst{3,4}, X. Hou\inst{13}, R. Iaria\inst{10}, G. Illiano\inst{7,1,2}, D. M. Russell\inst{14},  R. Sathyaprakash\inst{15}
          }

   \institute{Institute of Space Sciences (ICE, CSIC), Campus UAB, Carrer de Can Magrans s/n, E-08193 Barcelona, Spain
         \and
             Institut d'Estudis Espacials de Catalunya (IEEC), 08860 Castelldefels (Barcelona), Spain 
             \and
             Departamento de Física, Universidad de Santiago de Chile (USACH), Av. Víctor Jara 3493, Estación Central, Chile
             \and
             Center for Interdisciplinary Research in Astrophysics and Space Sciences (CIRAS), Universidad de Santiago de Chile.
             \and
             INAF/IASF Palermo, via Ugo La Malfa 153, I-90146 - Palermo, Italy 
             \and 
            Université Grenoble Alpes, CNRS, IPAG, F-38000 Grenoble, France
              \and
              INAF–Osservatorio Astronomico di Brera, Via Bianchi 46, I-23807 Merate (LC), Italy
              \and
              INAF Osservatorio Astronomico di Roma, via Frascati 33, 00078, Monteporzio Catone (Roma), Italy
              \and
             Universit\`a degli Studi di Cagliari, Dipartimento di Fisica, SP Monserrato-Sestu km 0.7, I-09042 Monserrato, Italy
              \and
            Universit\`a degli Studi di Palermo, Dipartimento di Fisica e Chimica, via Archirafi 36 - 90123 Palermo, Italy
             \and
            IRAP, CNRS, Université de Toulouse, OMP, CNES, 9 avenue du Colonel Roche, BP 44346, F-31028 Toulouse Cedex 4
            \and
            European Space Agency (ESA), European Space Astronomy Centre (ESAC), Camino Bajo del Castillo s/n, E-28692 Villanueva de la Cañada, Madrid, Spain
            \and
            Yunnan Observatories, Chinese Academy of Sciences, Kunming 650216, People’s Republic of China
            \and
            Center for Astrophysics and Space Science (CASS), New York University Abu Dhabi, PO Box 129188, Abu Dhabi, UAE
            \and
            Scuola Universitaria Superiore IUSS Pavia, Piazza della Vittoria 15, I-27100, Pavia, Italy
             }
  
   \date{XXX-XXX}
 
  \abstract{Accreting millisecond pulsars (AMSPs) and transitional millisecond pulsars (tMSPs) are neutron star low-mass X-ray binaries which can evolve into "recycled" radio millisecond pulsars. In both types of systems, X-ray pulsations have been detected during phases of X-ray activity when matter accretion through a disc is turned on. On the other hand, when accretion stops, and these systems enter the quiescent, low-luminosity X-ray state, only tMSPs become visible as radio pulsars. Despite several attempts, radio pulsations have never been detected in quiescent AMSPs, except for IGR\,J18245$-$2452. In this manuscript, we present the results of two observational campaigns performed on the AMSPs \igr and \maxij with the Five-hundred-meter Aperture Spherical Telescope (\fast) in L-band (1-1.5 GHz). Both sources have most likely been observed in quiescence, as suggested by the upper limits on their X-ray and optical flux obtained with \swift\ and the Las Cumbres Observatory, respectively. We have performed a deep search for coherent periodicities in radio but found no significant candidate signal, either at the known spin frequency of those sources or at other frequencies. Assuming a pulse duty cycle of 10\%, we derive upper limits on the pulsed radio flux density of 3.3~$\mu$Jy and 5.6 $\mu$Jy for \igr and \maxij , respectively, which are the most stringent limits so far for any known persistent AMSP. 

}
 

  \authorrunning{A. Marino}
  \titlerunning{\fast observations of \igr and \maxij in quiescence}
  \maketitle

\section{Introduction}
Accreting millisecond pulsars (AMSPs\footnote{ Please note that they are often referred to as accreting millisecond X-ray pulsars, AMXPs in short.}) and transitional millisecond pulsars (tMSPs) are classes of low-mass X-ray binaries (LMXBs) that harbour X-ray pulsars spinning at frequencies of about hundreds of Hz \citep[see][for reviews]{Campana2018, Patruno2021, DiSalvo2022, Papitto2022, DiSalvo2023}. These systems are transient, as they spend most of their life in a low-luminosity quiescent X-ray state ($L_X \lesssim 10^{33}$ erg s$^{-1}$), where matter accretion is mostly shut-off, interrupted by sporadic episodes of X-ray activity. In AMSPs, these episodes are typically called outbursts and are short, i.e., lasting from weeks to months \citep[see][for a summary of their outburst properties]{Marino2019a}, and bright, with peak X-ray luminosities $L_X \sim 10^{36}-10^{37}$ erg s$^{-1}$. tMSPs, on the other hand, can show longer, low luminosity "disc" or "intermediate" states, i.e., lingering at $L_X \sim 10^{33}-10^{34}$ erg s$^{-1}$ for decades characterized by bi-modal, multi-wavelength variability \citep[see, e.g.][]{deMartino2010,deMartino2013, Linares2014, Bogdanov2015, Jaodand2016, Baglio2023}. During these episodes of X-ray activity, coherent pulsations have been observed in both classes over a wide range of electromagnetic frequencies, including X-rays, optical and UV \citep{Ambrosino2017,Ambrosino2021}. The two classes however seem to behave differently during quiescence: while tMSPs exhibit coherent radio pulsations \citep{Archibald2009,Papitto2013,Stappers2014,Bassa2014}, searches for such pulsations in about one-fifth of the known AMSPs have consistently yielded non-detections \citep{Burgay2003,Iacolina2009,Iacolina2010,Patruno2017_1808,Sanna2018_16597,Gusinskaia2020,Li2026}, with the
most stringent flux density limits ranging between 2.9–5.7 $\mu$Jy \citep[in L-band,][]{Peng2025}. The only AMSP that effectively switches on as a radio pulsar in quiescence is IGR\,J18245$-$2452, which belongs to both classes \citep{Papitto2013}. The characteristic swinging between accretion-powered pulsar states and rotation-powered pulsar states observed in tMSPs can be considered an observational key to confirm the so-called "pulsar recycling scenario" \citep{Alpar1982}. According to this hypothesis, radio millisecond pulsars \citep[MSPs; see][for reviews]{Lorimer2008,Borghese2023} are descendants of accreting NSs in LMXBs binaries, having been spun-up by prolonged mass accretion from a low-mass companion star. 

Given their importance, considerable efforts have been devoted to identifying new tMSP candidates. On the one hand, these searches have targeted unidentified $\gamma$-ray sources - most likely belonging to the "spider" pulsars family \citep{Eichler1988,Roberts2013} - that exhibit a sub-luminous “disc” state similar to tMSPs, \citep[e.g.][]{Bogdanov2015_J1544,CotiZelati2019,Cotizelati2021,CotiZelati2024,Jaodand2021,Manca2025_tMSP,Gusinskaia2025,Kyer2025}. On the other hand, it remains unclear whether at least some AMSPs may switch on as radio pulsars in quiescence and thereby behave as tMSPs as well. Solving this conundrum would have repercussions on several aspects of accretion, binary and pulsar physics. Indeed, although both AMSPs and tMSPs can be considered progenitors of radio MSPs, it remains unclear whether they constitute distinct evolutionary classes or they represent different phases of a continuous evolutionary sequence. Additionally, detecting radio pulsar activation in quiescence would help explain the secular spin-down observed in some AMSPs \citep{DiSalvo2022}, and shed light on how the neutron star magnetosphere and the accretion flow interact at low mass-accretion rates.

Expanding the sample of quiescent AMSPs observed in radio, performing regular, high-cadence monitoring programs and/or taking advantage of more sensitive instruments are therefore crucial ways to address this open question. As the largest single-dish radio telescope in the world, no instrument is better suited to the latter quest than \fast. Unfortunately, besides Aql X$-$1, only two of the 26 sources in the AMSPs class \citep[see][for updated lists]{DiSalvo2023, Ng2024_atel}, i.e. \igr and \maxij, are visible from the \fast location. In this manuscript, we report on \fast observations performed on both targets. The manuscript is organized as follows. We present the target sources and their main properties in Sections \ref{ss:igrj}–\ref{ss:maxij}, describe the multi-band data reduction in Section \ref{sec:obs}, outline the analysis and radio pulsation searches in Section \ref{sec:data}, and discuss the results in the context of current literature and future prospects in Section \ref{sec:disc}.

\subsection{\igr}\label{ss:igrj}
\igr holds the record as the fastest-spinning NS among AMSPs with a spin frequency of $\sim$599 Hz \citep{Markwardt2004}. Since its discovery in 2005 \citep{Shaw2005}, the system displayed three other short outbursts: two in 2008 \citep{Lewis2010} and one in 2015 \citep{Patruno2017_J00291, DeFalco2017, Sanna2017}, with a duration of 14, 24 and 25 days, respectively. The orbital period is 2.46 hrs, while the distance of the system has been estimated to be 4.2$\pm$0.5\,kpc from the analysis of a photospheric radius expansion type-I X-ray burst \citep{DeFalco2017}. Optical observations of the system in quiescence suggest an irradiated companion star, attributed to the wind of a radio pulsar \citep{Davanzo2007}. Furthermore, an episode of intense optical flaring activity (e.g. \citealt{Baglio2017}) hinted at the presence of an accretion disc even at low X-ray luminosity, somewhat reminiscent of the sub-luminous activity observed in tMSPs \citep{Papitto2018, Kennedy2018}. However, to our knowledge, no search for radio emission in the system during quiescence has been conducted thus far. 
\subsection{\maxij}\label{ss:maxij}
 \maxij is an ultra-compact X-ray binary system \citep{ArmasPadilla2023} with an orbital period of only $\sim$61 mins \citep{Bult2022_atel_1957}. A constraint on the source distance of about 5$\pm$2 kpc was placed by comparing the observed optical magnitude of the companion star with the one expected for a late-K/early-M dwarf star \citep{Ravi2017}. The behaviour of the source is intriguing, as it shows frequent, i.e. four in about one year \citep{Sugimoto2015, Tanaka2016}, but short (i.e. $\sim$days) and faint outbursts, with peak luminosities of about 10$^{36}$ erg s$^{-1}$ or lower \citep{Matasanchez2017_1957, Beri2019}, the most recent one in May 2025 \citep{Illiano2025_1957, sanna2026}. Although \maxij has been known for over a decade, it was identified as a 3.2-ms AMSP only during an outburst in 2022 \citep{Sanna2022_1957}. 
 
\section{Observations \& Data Reduction}\label{sec:obs}

\subsection{FAST}
The \fast \sout{Telescope} \citep{Nan2011} is currently the largest and most sensitive single-dish radio telescope on Earth, making it the best instrument to probe faint radio emission. Two 50-min observations of \igr were performed with \fast on 2022 September 22 and October 23. \maxij was observed once on 2023 October 8 for 20 minutes (see Tables \ref{tab:fast-igrj}-\ref{tab:fast-maxij}). All observations were carried out in Tracking mode with the central beam of the L-band 19-beam receiver. Search-mode data were recorded with the ROACH-2 backend over a frequency range of 1 to 1.5 GHz. The 500-MHz band was sampled at a rate of 49.152\,$\mu$s into 4096 channels, corresponding to a channel width of 0.122 MHz. We scheduled the observations such that the pulsars were as close as possible to zenith at the \fast site and near inferior conjunction in their respective orbits based on the extrapolation of the orbital ephemerides of these systems (based on the orbital ephemerides from \citealt{Sanna2017} for \igr and \citealt{Sanna2022_1957} for \maxij). This strategy was adopted to minimize the probability of non-detection caused by eclipses of the radio signal by circumbinary material, which is more likely when the pulsar is near superior conjunction. 

\begin{table}[]
            \centering
            \caption{Journal of the multi-wavelength observations used in this work for \igr.}
            \label{tab:fast-igrj}
            \begin{tabular}{ccc}
            \hline\hline
            Obs. Date & Duration & ObsID \T \\
            YYYY-MM-DD & s & \B \\
            \hline
            \multicolumn{3}{c}{\fast} \T \\
            \hline
            2022-09-22 16:43:00 & 3000 & \T \\
            2022-10-23 14:39:00  & 3000 & \B \\
            \hline 
            \multicolumn{3}{c}{\swift/XRT} \T \\
            \hline
            2022-09-22 13:40:36 & 4900 & 00031253007 \T \\
            2022-10-23 07:47:35 & 4400 & 00031253008 \B \\
            \hline 
            \multicolumn{3}{c}{\it LCO} \T \\
            \hline
            2022-09-22 21:48:29  & 6100 \T  \\
            2022-10-14 05:39:56 & 300 \\
            2022-10-23 05:45:47 & 300 \\
            2022-10-26 07:40:42 & 300  \B \\
            \hline
            \hline 
            \end{tabular}   
        \end{table}

\begin{table}[]
            \centering
            \caption{Journal of the multi-wavelength observations used in this work for \maxij.}
            \label{tab:fast-maxij}
            \begin{tabular}{ccc}
            \hline\hline
            Obs. Date & Duration & ObsID \T \\
            YYYY-MM-DD & s & \B \\
            \hline
            \multicolumn{3}{c}{\fast} \T \\
            \hline
            2023-10-08 11:38:00 & 1200 & \T \B \\
            \hline 
            \multicolumn{3}{c}{\swift/XRT} \T \\
            \hline
            2023-10-01 13:49:55 & 1500 & 00033770033 \T \B \\
            \hline
            \hline 
            \end{tabular}   
        \end{table}

\subsection{Swift}
The sources were also observed within about one week of the \fast observations with the X-ray telescope \citep[XRT, e.g.][]{Burrows2005} on board the Neil Gehrels Swift Observatory (\swift) to check whether their X-ray fluxes were compatible with the systems being in quiescence. In particular, \igr was observed twice: once on 2022 September 22 for 4.9 ks and on 2022 October 23 for 4.4 ks. \maxij was observed on 2023 October 1 for 1.5 ks (as reported in Tables \ref{tab:fast-igrj}-\ref{tab:fast-maxij}). All the observations were performed with the XRT in Photon Counting (PC) mode. All the tools used for the X-ray data reduction are part of the \textsc{HEASOFT} software package (v. 6.32.1). The XRT observations were downloaded from the \textsc{HEASARC} public archive\footnote{\url{https://heasarc.gsfc.nasa.gov/docs/archive.html}} and processed using the standard XRT data reduction tools, \textsc{xrtpipeline} and \textsc{xrtproducts}. The latest version of CALDB available at the time these data were released was used. Once we ensured that the pile-up effect was negligible in all observations, we used circular regions of $\sim$20 arcseconds centred on the source's coordinates to extract photons from the source. An annular region of inner (outer) radius of 45$^{\prime \prime}$ (110$^{\prime \prime}$) was used to estimate the background counts.
        
\subsection{Las Cumbres Observatory}
\igr was first observed on 2022 September 22, one day after the first \fast observation, with the 1-m telescope of the Las Cumbres Observatory (LCO) network located at the Teide Observatory (Tenerife, Canary Islands). Observations lasted $\sim1.7$ hours, between 21:48:29.339 UTC and 23:30:59.671 UTC, and consisted of 16 consecutive pointings using the SDSS $i'$ filter (central wavelength $7480\AA$), each with an exposure time of 300s.
Images were taken under poor seeing conditions ($\sim 3-4^{\prime \prime}$). 

The source was observed again with the same filter and exposure time (300s) on October 14 (05:39:56.453 - 05:44:56.427 UTC), 20 (09:05:50.140 - 09:10:50.140 UTC), 23 (05:45:47.157 - 05:50:47.157 UTC) and 26 (07:40:42.466 - 07:45:42.466 UTC), using the 1-m telescope of the LCO network located at the McDonald Observatory (USA). Seeing conditions during these three observations remained below 22$^{\prime \prime}$. A summary of these observations is reported in Table \ref{tab:fast-igrj}. We performed image-reduction by subtracting an averaged bias frame and dividing by a normalized flat frame. Aperture photometry was then carried out on all the stars in the field using the {\tt daophot} tool within the Starlink distribution, adopting an aperture radius of approximately 1.5 times the FWHM of the stellar profiles. We performed flux calibration against seven isolated stars in the field with magnitudes tabulated in the APASS catalogue \citep{Henden2012}. We estimated the photometric upper limits from the magnitude of the faintest reliably detected source in the field, which has a magnitude uncertainty of $\sim0.3$ mag, corresponding to a signal-to-noise ratio of about 3. 

\maxij is monitored with LCO as part of an ongoing program targeting approximately 50 LMXBs \citep{Lewis2008}. As shown by \citet{sanna2026}, its quiescent optical counterpart is blended with a nearby source ($g'\sim20$ mag) located $\sim2^{\prime \prime}$ away, making reliable separation and photometry in quiescence impossible with our instrument configuration. Therefore, we do not discuss the quiescent optical counterpart of \maxij in this work.

\section{Data analysis}\label{sec:data}

Data analysis was performed using the \texttt{PRESTO}\footnote{\url{https://www.ascl.net/1107.017}} \citep{Ransom2002} pulsar analysis software. The \texttt{rfifind} tool was used to excise radio frequency interference (RFI) from the data. 

We searched the data for radio emission using several methods, which are described in the sections below. Because the dispersion measure (DM) of each source is unknown, this parameter had to be explored blindly in all searches. Along the line of sight (LoS) toward \igr, and assuming a distance of 4.2\,kpc, the NE2001 \citep{Cordes2002} and YMW16 \citep{Yao2017} Galactic electron density models predict DM values of approximately 130 and 165\dmu, respectively, with maximum Galactic DMs of 220 and 260\dmu. For \maxij, at the estimated distance of 5\,kpc, the NE2001 and YMW16 models predict DMs of $\sim$106 and 73\dmu, and maximum Galactic DMs of 138 and 107\dmu. Given the substantial uncertainties typically associated with DM predictions from Galactic electron density models, as well as the uncertainties in the source distances (see Sections \ref{ss:igrj}–\ref{ss:maxij}), we searched over trial DM values ranging from 0 to 400\,pc\,cm$^{-3}$ in each search method—well above the maximum Galactic DM predicted for both systems. To prevent intra-channel dispersion smearing from reducing pulse detectability, we used a DM step of 0.1\,pc\,cm$^{-3}$, resulting in a total of 4000 DM trials.

\subsection{Search for periodicity in \fast data}
\subsubsection{Brute-force folding using known ephemerides}\label{ss:brute-force-folding}
We first searched for pulsations by folding the cleaned data using the known X-ray ephemeris of each source (see Table \ref{tab:ephemer}) with the \texttt{PRESTO} \texttt{prepfold} routine. To take into account the uncertainties on the predicted ephemerides (notably due to the large errors of the orbital parameters in the timing models) at the epoch of the \fast observations, we enabled \texttt{prepfold} to search in the $P$-$\dot{P}$ parameter space around the instantaneous values predicted by the timing model. This effectively enabled a search in line-of-sight acceleration resulting from the orbital motion of the pulsar, $a_l = \dot{P}c/P$, where $c$ is the speed of light.  

We folded the data at each DM trial, allowing \texttt{prepfold} to perform for a narrow search around the trial DM, and inspected all diagnostic plots. Considering a threshold of S/N$>$6 for a detection, no significant pulsations were identified for either source. 

\subsubsection{Acceleration search} \label{sec:accelsearch}

Given the several-year gap between the epoch of our \fast\ observations and the epochs of the reference X-ray timing solutions, as well as their limited precision, the absence of a detection in the brute-force folding search may be due to the original ephemerides no longer accurately describing the systems, for instance, as a result of orbital evolution. We therefore conducted an independent search for periodic signals using the \texttt{PRESTO} \texttt{accelsearch} routine, performing Fourier-domain searches for periodicities across all 4000 dedispersed time series generated for each \fast\ observation. 

Because the duration of all \fast\ observations, $T_{\rm obs}$, corresponds to about 30\% of the system's orbital period ($T_{\rm obs} \approx 0.3P_B$), significant orbital jerk is expected. Although the \texttt{PRESTO} \texttt{accelsearch} routine includes a jerk search implementation \citep{Andersen2018}, substantial spreading of signal power over multiple Fourier bins persists when $T_{\rm obs} \gtrsim 0.15P_B$ due to uncorrected higher-order acceleration effects \citep{Andersen2018}. Since there is no quantitative prescription for the loss of Fourier-domain signal power when $T_{\rm obs}/P_B \gtrsim 0.15$, we performed searches using both the full observation and segmented time series \citep[as in, e.g.][]{Sathyaprakash2019}. This combined strategy balances signal coherence and computational efficiency, maintaining sensitivity to pulsars in compact binary systems.

\begin{enumerate}
    \item First, we carried out a Fourier-domain jerk search with \texttt{accelsearch} over the full-length time series to maximize sensitivity to weak signals; to limit computational cost, this search was restricted to a narrow frequency range (a few tens of Hz) centred on the known spin frequency of each pulsar. The maximum acceleration and jerk along the line of sight that were searched are 20\,m\,s$^{-2}$ and 1.2$\times10^{-2}$m\,s$^{-3}$ for \igr, and 24\,m\,s$^{-2}$ and 4.3$\times10^{-2}$m\,s$^{-3}$ for \maxij. 

    \item Second, we divided each time series into two equal segments such that $T_{\rm obs}/P_B \sim 0.15$, and searched each segment with \texttt{accelsearch}. Segmenting the data ensures the validity of the linear-acceleration approximation and preserves signal coherence. Since the radiometer sensitivity scales as 1/$\sqrt{T_{\rm obs}}$, segmenting the observation into two equal-duration data sets results in an approximate 60\% reduction in sensitivity. However, as noted previously, it remains uncertain how this reduction compares to the loss of signal power expected under the constant-acceleration assumption in the regime $T_{\rm obs}/P_B \gtrsim 0.15$ (see e.g., \citealt{Andersen2018}). We searched the segmented time series to higher acceleration and jerk values: 26\,m\,s$^{-2}$ and 2.9$\times10^{-2}$m\,s$^{-3}$ for \igr, and 117\,m\,s$^{-2}$ and 20.5$\times10^{-2}$m\,s$^{-3}$ for \maxij. To limit computational costs, we set the maximum number of harmonics being summed incoherently with \texttt{accelsearch} to two.

\end{enumerate}

Fourier-domain detections, characterized by an equivalent Gaussian significance $\sigma_F$\footnote{Here $\sigma_F$ denotes the equivalent Gaussian significance derived from the false-alarm probability of the measured Fourier power, computed assuming the appropriate $\chi^2$ distribution and accounting for the number of harmonics summed; see \citet{Ransom2002} for details.}, greater than 1.5 across all dispersion-measure (DM) trials were sifted to produce final candidate lists. During the sifting stage, detections consistent in spin frequency across multiple DMs were grouped together, and for each group the DM yielding the highest significance was retained while harmonically related and low-significance duplicates were rejected. Candidates with the highest significances ($\sigma_F > 5$) were subsequently folded using \texttt{prepfold}. Visual inspection of the resulting folded profiles did not reveal any convincing pulsar candidates.

\subsubsection{Brute-force search in the time of ascending node}\label{ss:spider_twister}

We searched the dedispersed time series using the \texttt{SPIDER\_TWISTER} package\footnote{\url{https://github.com/alex88ridolfi/SPIDER_TWISTER}} to correct for possible drift in the pulsar's time of passage through the ascending node, $T_{\rm asc}$. The software performs a grid search over $T_{\rm asc}$, folding the data using the reference timing ephemeris for a range of trial values and identifying the one that maximizes the S/N of the folded profile. This method efficiently recovers pulsations in “spider” systems whose orbital periods evolve stochastically and can render older ephemerides inaccurate \citep{rft+16}. For each source, we used a number of $T_{\rm asc}$ trials twice that of the minimum required by Eq.~5 of \citet{rft+16}, ensuring complete coverage of plausible orbital-phase offsets.

We visually inspected the $\sim$4000 diagnostic \texttt{prepfold} plots produced at each trial DM for every observation of both sources. While no promising signal was identified for \maxij, a small number of tentative candidates were found for \igr in both observations and warranted further scrutiny. We subsequently refolded the raw data at the candidate DMs following the procedure described in Section~\ref{ss:brute-force-folding}, and performed a finer search in the vicinity of each candidate DM. In addition, we explored several combinations of \texttt{prepfold} folding parameters (e.g., number of profile bins, subintegrations, and frequency subbands) to better assess the robustness of the putative detections, as false positives arising from random noise fluctuations are known to be highly sensitive to binning choices. The resulting diagnostic plots were re-examined, but no convincing signals were recovered, leading us to conclude that the initially selected candidates are consistent with false positives. This interpretation is further supported by the following considerations: (i) none of the candidates exhibits a coherent detection across neighbouring or consecutive DM trials, whereas a genuine pulsar signal would typically persist across several ($\gtrsim$2--3) adjacent DM values; (ii) the apparent signals vanish when alternative \texttt{prepfold} parameter settings are adopted; (iii) the candidates are not statistically significant relative to the distribution of folding statistics obtained from the full set of \texttt{SPIDER\_TWISTER} trials; and (iv) no candidate is detected at a consistent DM in both observations.

\subsubsection{Single-pulse search}
We also searched for sporadic radio emission in the form of single pulses from both \igr and \maxij. Although no AMSP has yet exhibited confirmed single-pulse radio emission, such signals could plausibly arise from transient magnetospheric activity or binary interaction processes. In tMSPs, rapid switches between accretion- and rotation-powered states are known to produce abrupt changes in radio output and intermittent coherent emission \citep{Archibald2009,Papitto2013,Bassa2014}. Sporadic bursts could also originate from magnetospheric reconnection or pulsar wind–companion shock emission during quiescent phases. Furthermore, some rotation-powered MSPs emit short-duration giant pulses, as observed in PSR~B1937+21 \citep{cstt96}, suggesting that similar emission mechanisms might operate intermittently in AMSPs once accretion subsides. To probe possible sporadic or burst-like emission in the \igr and \maxij systems, we searched all dedispersed time series (over the full 0–400\,pc\,cm$^{-3}$ DM range) with \texttt{PRESTO}’s \texttt{single\_pulse\_search.py} for dispersed pulses using boxcar widths of 0.1–100\,ms. No astrophysical single-pulse candidates with S/N$>$6 were identified.

\begin{table}
\centering
\caption{Orbital parameters and spin frequency of \igr and \maxij used in this work. See, for more details, the cited papers.}
\begin{tabular}{l p{3 cm}  | p{3 cm} }
\hline
\hline
Parameters &  {\igr} & {\maxij} \\
\hline
R.A. (J2000) & {$00^h29^m3.05^s$} & $19^h56^m39.11^s$ \\
DEC (J2000) & {$+59^\circ34^m18.93^s$} & $+03^\circ26' 43.7\arcsec$ \\
$P_{\rm orb}$ (d) & {0.1023620(2)} & 0.0422807(7) \\
$x$ (lt-s) & {0.0649905(24)} & 0.013796(25) \\
$T_{\rm ASC}$ (MJD) & {57231.437581(3)} & 59749.633146(18) \\
$e$ & {<$2\times 10^{-4}$} & $ < 1.4 \times 10^{-2}$\\
$\nu_0$ (Hz) &598.8921309(2) & 313.64374049(22) \\
$\dot{\nu}_0$ (Hz/s) &$-4.1(1.1)\times 10^{-15}$ & $-2.2(0.4)\times 10^{-14}$ \\
\hline
Ref. & \cite{Sanna2017, Patruno2017_J00291} & \cite{Sanna2022_1957}, \cite{sanna2026} \\
\hline
\hline
\end{tabular}
\label{tab:ephemer}
\begin{tablenotes}
\item {$P_{\rm orb}$: Orbital period in days; $x$: orbital projected semi-major axis, $T_{\rm ASC}$: time of passage of the NS at the ascending node; $e$: orbital eccentricity; $\nu_0$: spin frequency;  $\dot{\nu}_0$: spin frequency derivative.}
\end{tablenotes}
\end{table}

\subsection{Upper limits on the radio pulsed emission}
In the absence of detected pulsations, we placed upper limits on the minimum detectable mean radio flux density $S_{\rm min}$ from both sources. We used the radiometer equation \citep{dtws85}: 
\begin{equation} \label{eq:radiometer}
    S_{\rm min}=\beta \frac{S/N_{\rm min} \left( T_{\rm rec} + T_{\rm sky}\right)}{\mathcal{E}\, G\sqrt{N_{\rm p}T_{\rm obs}\Delta f_{\rm eff}}} \sqrt{\frac{\delta}{1-\delta}}~,
\end{equation}
where $S/N_{\rm min}$ is the threshold signal-to-noise ratio of our search, $\beta$ is the digital correction factor, $T_{\rm rec}$ is the receiver temperature, $T_{\rm sky}$ is the sky temperature at 1.25\,GHz\footnote{Sky temperatures at the pulsar positions were estimated by extrapolating the 408-MHz Galactic radio continuum emission \cite{hss+82,rdb+15} to 1250\,MHz assuming a spectral index of –2.85 \citep{dbc+19}.}  
,$\mathcal{E}(\delta)$ is the search efficiency (see explanation below) of the Fourier-based search as a function of pulse duty cycle $\delta=W/P$, $G$ is the telescope gain, $N_{\rm p}$ is the number of polarization channels summed, $T_{\rm obs}$ is the integration time, $\Delta f_{\rm eff}$ is the effective observation bandwidth. Here we used: $\beta$=1.1, $T_{\rm rec}$=25.7 K and $G$=16.7 K\,Jy$^{-1}$ \citep[see, for references on the assumed values][]{Jiang2020, Qian2020}, $N_p$=2, $\Delta f_{\rm eff}$\,=\,400\,MHz (RFI masking of 20\% of the full bandwidth). 

As discussed by \citet{yu18} and \citet{mbs20}, the $S/N_{\rm min}$ term in the original radiometer equation \citep{dtws85} corresponds to the true upper detection limit applicable to phase-coherent folding searches. In contrast, our Fourier-based incoherent search adopts a Gaussian-equivalent significance expressed in units of $\sigma$. This choice leads to an overestimation of the predicted search sensitivity, particularly for pulse profiles with substantial harmonic content (i.e., small $\delta$).

Following the formalism of \citet{mbs20}, we correct for this discrepancy by introducing a search efficiency factor, $\mathcal{E}(\delta)$, into the radiometer equation. This factor accounts for the incoherent summation of up to two harmonics in the FFT-based search (see Section~\ref{sec:accelsearch}). For representative pulse duty cycles of $\delta = 1\%, 10\%,$ and $50\%$, the corresponding efficiencies are $\mathcal{E}(\delta) = 0.25, 0.73,$ and $0.93$, respectively. See \citep{mbs20} for more details. 

\begin{figure}
\centering
\includegraphics[scale=0.7]{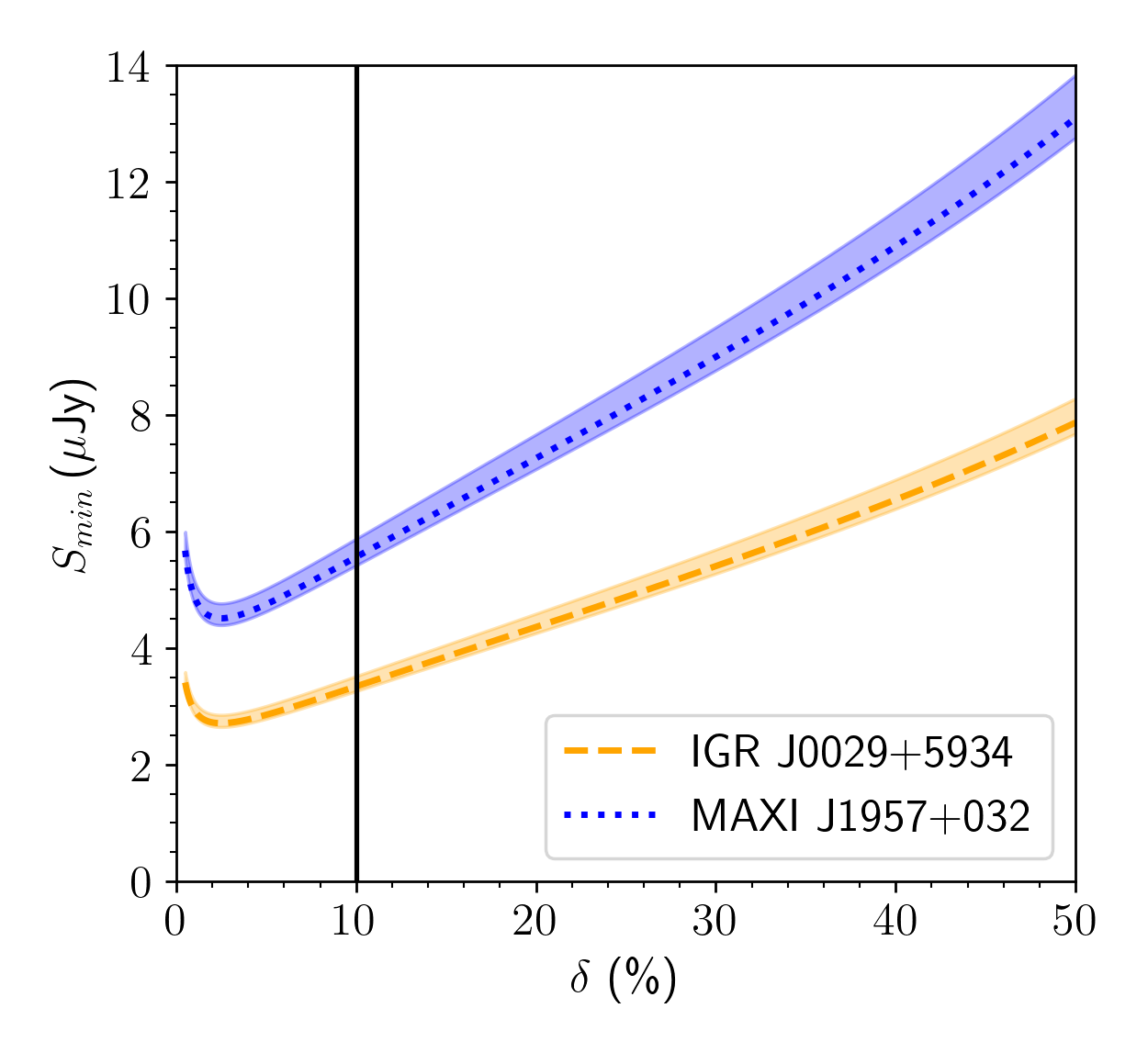}
\caption{Minimum detectable pulsed flux density $S_{\rm min}$ computed using Eq~\ref{eq:radiometer} in our FAST data as a function of pulse duty cycle $\delta$ for \igr\, (orange, dashed line) and \maxij (blue, dotted line) at the centre frequency (1250\,MHz) of the receiver band. The shaded areas represent the range of $S_{\rm min}$ when considering the varying sky temperature across the receiver. The assumed typical MSP duty cycle of 10\% is indicated as a vertical black line \citep{Manchester2005}.} 
\label{Fig:Sminvsd}
\end{figure}

Assuming a detection threshold of $S/N_{\rm min} = 5$ and incorporating the efficiency correction $\mathcal{E}(\delta)$, Figure~\ref{Fig:Sminvsd} illustrates the resulting dependence of the minimum detectable flux density, $S_{\rm min}$, on duty cycle $\delta$ for both sources.
The sample of known radio MSPs has a median pulse duty cycle $\delta$ at L-band of 10\% \citep{Manchester2005}. Assuming this value for $\delta$, we derived stringent upper limits of $S_{\rm min}$ = 3.3 $\mu$Jy for \igr and $S_{\rm min}$ = 5.6 $\mu$Jy for \maxij.

\subsection{Upper limits on the optical and  X-ray emission}

In none of the \swift /XRT observations and the LCO optical images used in this work, the source, either \igr or \maxij, was detected. For the optical emission of \igr, we estimate a $3\sigma$ upper limit in the range 19.9 - 21.6 mag (AB system). Regarding the X-ray emission, we initially used the command \texttt{sosta} (source statistics) within \textsc{Ximage}  to count the number of events within the selected source region and, by taking into account vignetting, exposure, point spread function and background counts, estimate an upper limit on the count rates, The estimated upper limits on the count rates were then translated into fluxes in the 0.5-10 keV energy range employing the web version of \textsc{PIMMS} (Portable Interactive Multi-Mission Simulator\footnote{This tool uses an assumed simple, e.g. blackbody or power-law, spectral model and folds it through the response matrix of an X-ray instrument to predict expected count rates or fluxes, see \url{https://heasarc.gsfc.nasa.gov/cgi-bin/Tools/w3pimms/w3pimms.pl}.} for more details.). To perform the conversion, we assumed a power-law spectrum of $\Gamma\sim$2.0 and $N_{\rm H}$ of 4$\times$10$^{21}$\,cm$^{-2}$ for both \igr and \maxij, as both systems have been reported to have similar hydrogen column density $N_H$ values \citep{DeFalco2017, Beri2019}. It should be noted that assuming a power-law spectral shape is a conservative approach. In fact, if we had assumed a cold blackbody, e.g. with a temperature $kT_{\rm bb}\sim0.1-0.2$\,keV \citep[typical for quiescent NS Low Mass X-ray Binaries, e.g.][]{Degenaar2011, Servillat2012, Marino2018}, would result in even more constraining upper limits. Finally, we used the lower limits on the distances, i.e., 3.7 kpc for \igr \citep{DeFalco2017} and 3 kpc for \maxij \citep{Ravi2017}, to estimate the upper limits on the X-ray luminosity. We found that in all pointings, the sources likely had a luminosity $\lesssim$4-6$\times$10$^{32}$\,erg\,s$^{-1}$. These results are summarised in Table \ref{tab:x-rays-ul}. Therefore, irrespective of the assumed spectral shape, our results indicate that both were most likely in a quiescent state at the time of the \fast observations.

\begin{table}[]
            \centering
            \caption{X-ray count rate and luminosity upper limits (confidence level: 2.706 $\sigma$).}
            \label{tab:x-rays-ul}
            \begin{tabular}{ccccc}
            \hline\hline
            Source & ObsID & count rate & $L_{X}$ (0.5-10\,keV) \T \B\\
            & & ($\times10^{-3}$\,cts/s) & ($\times 10^{32}$ erg s$^{-1}$) \T \B\\
            \hline
            \multirow{2}{*}{J00291} & 00031253007  & $<$4.5 & $<$5.5 \T\\
                                  & 00031253008  & $<$3.5 & $<$4.2 \B\\
            J1957                & 00033770033  & $<$6.7 & $<$4.2 \T \B\\ 
            
            \hline 
            \end{tabular}
            
        \end{table}

\section{Discussion}\label{sec:disc}
We have carried out the deepest search to date for radio pulsations from the AMSPs \igr\ and \maxij\ during their X-ray quiescent phases using \fast. For both sources, no definitive pulsed signal was detected. For \maxij, this result confirms what was recently obtained with four different \fast observations performed in 2024 by \cite{Li2026}. In \igr, hundreds of marginal candidates have been identified (see Section \ref{ss:spider_twister}), but none of them can be reliably excluded as a false positive. These results align with the history of unsuccessful searches for radio pulsations in quiescent AMSPs, which we summarise in Table \ref{tab:radio-puls} and Fig. \ref{Fig:plot-upper-limits}. For all those sources, we have estimated the pseudo-luminosity $L_{\rm pseudo}$ by multiplying the measured radio flux density $S_\nu$ (or the upper limit on it) by the distance $d^2$, i.e. $L_{\rm pseudo}=S_\nu \times d^2$. It is noteworthy that the lack of radio pulse detections in quiescent AMSPs is in tension with several (indirect) observational clues suggesting that a radio pulsar should be active during quiescence. In particular: i) in eight AMSPs, the optical emission from the companion star showed clear modulation at the orbital period, possibly due to the star's surface being irradiated by the electromagnetic particle wind of a radio MSP\footnote{Although we note that even if this scenario holds, it does not necessarily imply detectable coherent radio pulsations at the time of the observations, since the pulsar wind can be present without active radio emission.\citep{Burderi2003,Campana2004, Davanzo2007, Davanzo2009}; ii) the discovery of optical and ultraviolet pulsations in SAX J1808.4$-$3658 during both the early and late stages of its outburst in 2019 could also be explained by invoking the reactivation of a rotation-powered mechanism \citep{Ambrosino2021}; iii) a gamma-ray counterpart has been proposed for SAX J1808.4$-$3658 \citep{deOnaWilhelmi2016}, which would make the source similar to other "spider" pulsars \citep{Roberts2013}.}

In the following, we discuss the upper limits obtained in this work within the emerging pattern of non-radio pulsating quiescent AMSPs and the elusive origin of such. 
\subsection{Possible scenarios for the non-detection of radio pulsations.}
A simple scenario that could account for the absence of radio pulsations in our sources is the presence of residual accretion and/or of a disc at the time of the radio observations, since even in tMSPs, radio pulsations are not observed during "disc states", with the most stringent upper limit to date being $\sim$2 $\mu$Jy with \fast \citep{Baglio2023}. In order to check whether our sources were in quiescence, we performed X-rays and optical observing campaigns within a few days of our \fast observations. The stringent upper limits obtained on their X-ray luminosity of 10$^{33}$ \ergs with \swift/XRT are lower than the typical X-ray luminosity of tMSPs in "disc states" \citep[10$^{33}$-10$^{34}$\,erg\,s$^{-1}$ , see e.g.][]{Papitto2022}, excluding that this type of activity was ongoing at the time of our \fast visits. Additionally, our optical upper limits of $i'>$19.9 - 21.6 mag (AB system) for \igr are consistent with the magnitudes reported in previous studies in quiescence \citep{Davanzo2007, Jonker2008, Baglio2017}, and are $\sim 3$ magnitudes fainter than the expected optical magnitude of the source in outburst ($i'\sim 17.2$ mag near the peak; \citealt{Lewis2010}). However, we note that \igr has displayed flaring activity in the past at magnitudes of about $i'\sim 20.7$ \citep{Baglio2017},  mostly consistent with or below the upper limits obtained in this work. Since such flaring was attributed to the presence of an accretion disc, we cannot entirely rule out the possibility of a faint, residual accretion disc at the time of our observations.

\begin{table*}
\centering
\caption{\bf Summary of radio detections/non-detections of AMSPs and tMSPs}
\begin{tabular}{l c c c c c c c}
\hline
\hline
Source &	$P_{\rm orb}$	& $\nu_{0}$ &	$d^{\dagger}$ & $S_\nu$ & Frequency range	&	Type & Ref.	\T \\
& (hr) & (Hz) & (kpc) & ($\mu$Jy) & (GHz) & \B \\
\hline
\igr &	2.46	& 599	& 4.2$\pm$0.5	& $<$3.3 & 1.0-1.5 &  AMSP & This work \T \B \\
\hline
\multirow{2}{*}{\maxij} &	\multirow{2}{*}{1.01}	& \multirow{2}{*}{313}	& \multirow{2}{*}{5$\pm$2} &		$<$5.6 	& \multirow{2}{*}{1.0-1.5} & \multirow{2}{*}{AMSP} & This work \T \\
 & & & & $<$12.3 & & & L26 \B \\
 \hline
SAX J1808.4$-$3658	& 2.01	& 401 &	3.3$\pm$0.1	&	$<$30 & 1.2-2.8 & AMSP & P17 \T \B \\
\hline
XTE J1751$-$305 & 0.71 &	435	& $<$8.5 & $<$31	&	7.9-9.0	& AMSP & I10 \T \B \\
\hline
XTE J0929$-$314 &	0.73	& 185 &	& $<$26 &	7.9-9.0 & AMSP & I09 \T \B \\ 
\hline
IGR J17591$-$2342 &	8.80	& 527	& $<$10	&	$<$26	&	1.7-2.6 & AMSP & G20 \T \B \\
\hline
PSR J1023+0038	& 4.75 &	592	& 1.3$\pm$0.1	&	750$\pm$80	& 1.6-2.4 & tMSP & A09 \T \B \\
\hline
XSS J12270$-$4859 & 6.91 & 593 &	1.4$\pm$0.1 & 660$\pm$20 & 0.1-0.9 & tMSP &  R15 \T \B \\
\hline
\multirow{4}{*}{IGR J18245$-$2453$^\ddag$} & \multirow{4}{*}{11.03} &	\multirow{4}{*}{254}	&  \multirow{4}{*}{5.5} & 60$\pm$30 & \multirow{4}{*}{1.2-2.8} & \multirow{4}{*}{AMSP, tMSP} & \multirow{4}{*}{P13} \T \\
& & & & 10$\pm$5 \\
& & & & 20$\pm$10 \\
& & & & 50$\pm$30 \B \\
\hline
XTE J1814$-$338 & 4.27 & 314 & 8.0$\pm$1.6 & $<$25.0 & 7.9-9.0 & AMSP & I10 \T \B \\
\hline
IGR J16597$-$3704 & 0.77 & 105 & 9.1 & $<$50.0 & 1.0-1.8 & AMSP & S18 \T \B \\	
\hline
Aql X$-$1 & 18.95	& 550 & 5.2$\pm$0.7 & $<$2.9 &  1.0-1.5 & Intermittent AMSP & P25 \T \B \\
\hline
\hline
\end{tabular}
\label{tab:radio-puls}
\begin{tablenotes}
\item {$P_{\rm orb}$: Orbital period; $\nu_0$: spin frequency; $d_{\rm kpc}$: distance \sout{in kpc}; $S_\nu$: radio flux density; $^\dagger$: we refer to \cite{Marino2019a} for a comprehensive list of references on the distance values reported here. $^\ddag$: Here we only report the detections of IGR J18245$-$2453, while we refer to \cite{Papitto2013}, Table 4 for a more comprehensive overview of all the radio measurements (including non-detections in quiescence) of the source;  References: A09=\cite{Archibald2009}, G20= \cite{Gusinskaia2020}, I09=\cite{Iacolina2009}, I10=\cite{Iacolina2010}, L26=\cite{Li2026}, P13=\cite{Papitto2013}, P17=\cite{Patruno2017_1808}, P25=\cite{Peng2025}, R15=\cite{Roy2015}, S18=\cite{Sanna2018_16597}}
\end{tablenotes}
\end{table*}

\begin{figure*}
\centering
\includegraphics[scale=0.47]{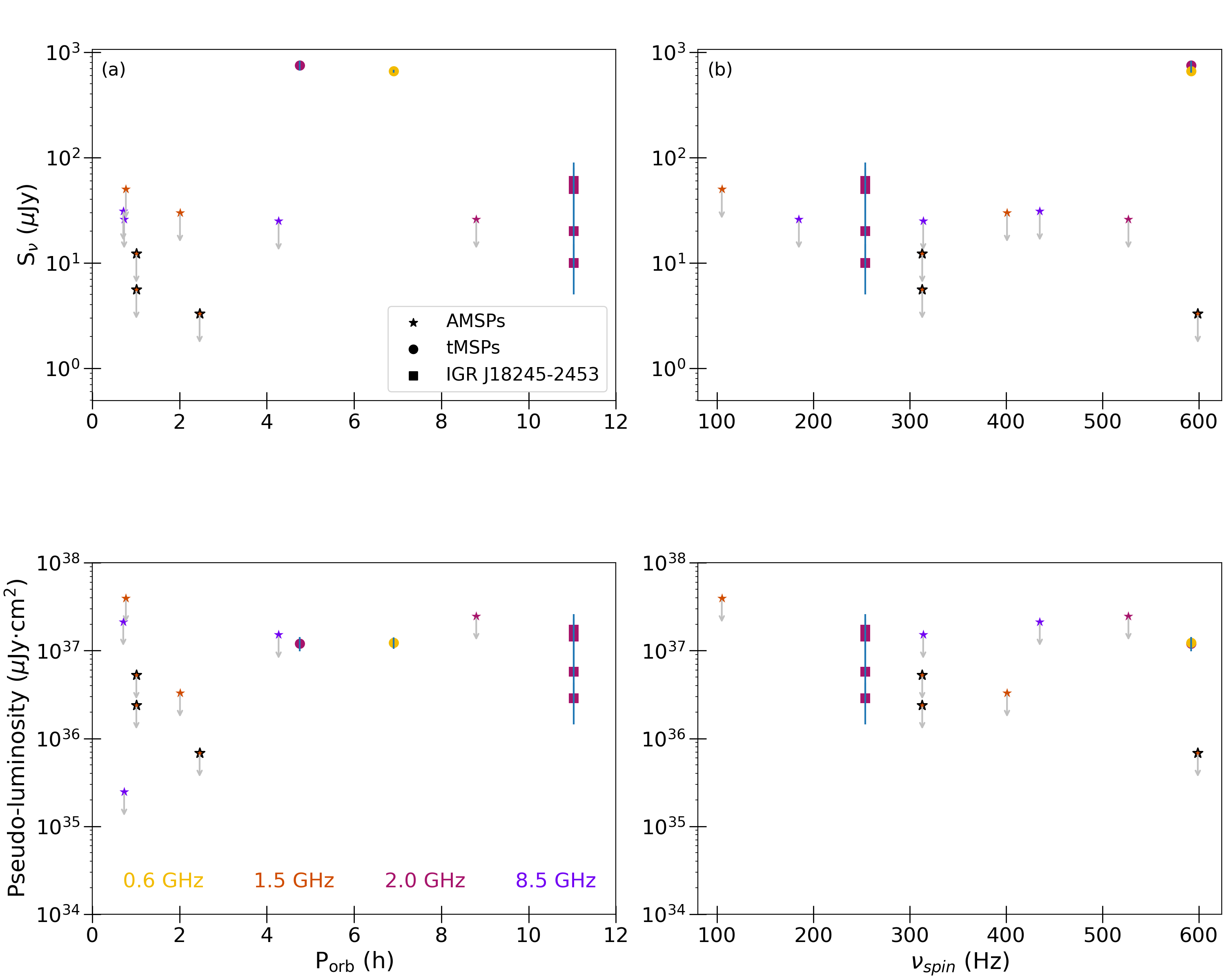}
\caption{Summary of radio detections/non-detections of AMSPs and tMSPs. In the top panels, we plot the measured value of $S_\nu$ vs. the orbital period (a) and spin frequency (b), while in the bottom panels, the pseudo-luminosity is displayed with respect to the same variable (c-d). Sources belonging to different classes are highlighted with different marker styles, while different colors are used to specify the frequency at which the radio observation was carried out. The sources studied in this paper are marked with black edges.} 
\label{Fig:plot-upper-limits}
\end{figure*}

Assuming that these sources were indeed in quiescence at the time of our observing campaigns, a simple scenario accounting for the radio non-detection of \igr and \maxij would be that they are exceptionally radio faint. For instance, the "ultra-faint" MSP PSR J0318.1+0252 discovered by \fast in 2018 - one of the faintest MSPs ever detected in radio - would have an observed flux density below those upper limits, if it were located at a distance comparable to the sources in our study \citep{Wang2021_J0318}. Furthermore, we note that the pseudo-luminosities measured for the three known tMSPs (see Fig. \ref{Fig:plot-upper-limits}) are largely consistent with the upper limits established for most AMSPs (excluding the sources analysed in this work), suggesting that these objects are comparatively faint in the radio band.

Alternatively, we consider the possibility that external factors may have hampered the detection of radio pulsations from our targets. For instance, the pulsar beam may have been unfavourably viewed, with a $\sim$66\% probability that it did not intersect our line of sight \citep{Iacolina2009}. Eclipses by the companion star could also contribute to the non-detection; however, this effect is expected to be minimal in our case, as the observations were timed to coincide with the pulsar’s inferior conjunction. Even in the presence of significant orbital evolution between outbursts, as sometimes observed in AMSPs \citep[e.g.][]{Sanna2017_1808,Illiano2023b}, any resulting shift in the time of passage would be negligible, i.e. within a few percent of the orbital period.

Finally, free-free scattering from material surrounding the pulsar may also account for the apparent lack of pulsations in our targets \citep[see][for a discussion]{Burgay2003}. This effect is known to alter drastically the pulsation detectability, even from orbit to orbit \citep[see, e.g.][for examples in tMSPs and “spider” pulsars]{Papitto2013,Zic2024}. In AMSPs, such a scenario is particularly plausible if a so-called "radio-ejection" mechanism \citep[consisting in the expulsion of transferred material by the pulsar wind radiation pressure, e.g.][]{Burderi2001} is active, as supported by recent general relativistic magneto-hydrodynamic (GRMHD) simulations \citep{Parfrey2017, Parfrey2024} and indirectly evidenced by their occasionally fast orbital expansion \citep{DiSalvo2008} and low average X-ray luminosity \citep{Marino2017, Marino2019a}. 

Interestingly, since less material is required to enshroud more compact systems, this effect is expected to anti-correlate with the orbital period \citep{Burgay2003}. A tentative trend is indeed suggested by archival radio searches in AMSPs (Fig.~\ref{Fig:plot-upper-limits} and Table~\ref{tab:radio-puls}): tMSPs typically have orbital periods longer than $\sim$4 hr, whereas AMSPs with reported (but unsuccessful) radio pulsation searches generally lie at shorter periods. In the future, searching for radio pulsations at higher frequencies, i.e., where the optical depth for free-free absorption is expected to be lower, could be a promising avenue to mitigate the impact of intra-binary material\footnote{However, we note that past searches at 5–8.5 GHz in four AMSPs were unsuccessful \citep{Iacolina2010}, possibly because these sources are intrinsically fainter at higher frequencies \cite[see][and references therein]{Burderi2003}.} Additionally, regular monitoring of these sources, even with less sensitive radio telescopes than \fast, would be crucial to check whether the presence of intra-binary material may make the detection of radio pulsations possible only sporadically.

\subsection{Do AMSPs simply not transition to the radio pulsar state in quiescence?}
Even considering that many factors may have negatively influenced our ability to detect pulsations from our target sources, the fact that 9 out of 10 AMSPs observed in the radio band during quiescence (see Table \ref{tab:radio-puls}) do not show pulsations begs to consider the alternative scenario where these sources may simply not switch to active radio pulsars when accretion stops. As a consequence, the secular spin-down observed in several AMSPs that displayed more than one outburst\sout{s} could not be explained with radio pulsar-like behaviour in quiescence, suggesting that other mechanisms, such as, e.g., residual torques on a cold disc or a weak propeller phase, have to be responsible \citep[see][and references therein]{DiSalvo2022}. 

If AMSPs and tMSPs are indeed distinct classes, the physical origin of this dichotomy remains unclear, whether it arises from differences in the long-term mass-accretion rate, the  nature of the companion star, and/or the microphysics of the coherent radio emission mechanism itself, which remains poorly understood. One distinction between the two classes seems to be the long-term behavior of their X-ray emission; the prolonged, sub-luminous X-ray active states observed in several candidate or confirmed tMSPs \citep[e.g.][]{Bogdanov2015, Cotizelati2021, Baglio2023} are indeed strikingly distinct from the shorter but significantly brighter outbursts typically observed in AMSPs, suggesting differences in the physical conditions in the accretion flows among the two classes. However, we note that this distinction is not unambiguous considering that: i) the mere existence of IGR J18245$-$2452 as a tMSP that underwent a single, bright AMSP outburst \citep{Papitto2013} proves that some sources can exhibit both types of X-ray activity; ii) "disc states" \emph{à la} tMSPs are intrinsically faint, with X-ray luminosities of the order 10$^{34}$ erg/s \citep{Papitto2022} and might thus be challenging to catch with all-sky X-ray monitors; iii) faint X-ray activity episodes, i.e., with X-ray luminosities between 10$^{34}$ and 10$^{35}$ erg s$^{-1}$, have been directly observed from some AMSPs \citep[e.g.][]{Wijnands2009, ArmasPadilla2013,Bult2019,Ng2021,Illiano2025_17511}. Sources exhibiting faint outbursts, known as very faint X-ray transients (VFXTs; \citealt{Muno2005}), have an uncertain physical origin \citep[e.g.][]{Heinke2015,Bahramian2021}. Given their similarities with tMSPs, some VFXTs have been proposed to host tMSPs in an X-ray active state \citep{Heinke2015}, making systems identified as both VFXTs and AMSPs promising candidates to activate as radio pulsars in quiescence. 

\section{Conclusions}
In this manuscript, we have conducted one of the deepest searches for radio pulsations in L-band ever performed for quiescent AMSPs, using the unprecedented capabilities of \emph{FAST}. The target sources, i.e., \igr and \maxij, along with the intermittent AMSP Aql X$-$1, are the only AMSPs known so far that are visible at the telescope site. Nevertheless, neither targets were detected as a radio pulsator. Using the radiometer equation and assuming a pulse duty cycle of 10\%, we have set a stringent upper limit of $\sim$3-6 $\mu$Jy on the pulsed radio emission from both these sources, the deepest \sout{up} to date for any AMSP. These limits are consistent with what was reported by \cite{Peng2025} for the intermittent AMSP Aql X$-$1, also observed with \fast. While \fast is not suited to observe the majority of AMSPs, which are predominantly located in the southern hemisphere, high-cadence monitoring with existing, even less powerful, telescopes remains a feasible avenue considering that pulse detectability may vary from orbit to orbit. Observations with future facilities of comparable or improved sensitivity, such as the Square Kilometre Array (SKA), will further help provide more definitive answers on whether some or all known AMSPs can turn on as radio pulsators during quiescence.

\begin{acknowledgements}
We thank the anonymous referee for their careful revision of this manuscript, which allowed us to significantly improve the quality of this manuscript. We thank B. Cenko and the \swift\ duty scientists and science planners for making the \swift\ ToO observations possible. AM acknowledges support from the Fund Vera Rubin/Chile 2024, under the project DIA 1736 "Silent black holes around red supergiants". AM, EP, FCZ and NR are supported by the H2020 ERC Consolidator Grant “MAGNESIA” under grant agreement No. 817661 (PI: Rea) and the National Spanish grant PID2023-153099NA-I00 (PI: Coti Zelati). EP is supported by a Juan de la Cierva fellowship (JDC2022-049957-I). FCZ is supported by a Ram\'on y Cajal fellowship (grant agreement RYC2021-030888-I). FCZ thanks Alessandro Ridolfi for sharing a version of his code for calculating the visibility of binary pulsars at specific orbital phases. This work was also partially supported by the program Unidad de Excelencia Maria de Maeztu CEX2020-001058-M, and by the PHAROS COST Action (No. CA16214). MCB acknowledges support from the INAF-Astrofit fellowship. AP was supported by INAF Research Grant RF2022 (FANS, PI: Papitto) and RF2024 (PULSE-X, PI: Papitto), the Italian Ministry of University and Research (PRIN MUR 2020, Grant 2020BRP57Z, GEMS, PI: Astone), and Fondazione
Cariplo/Cassa Depositi e Prestiti (SPES, Grant 2023-2560, PI: Papitto). CME acknowledges support from ANID/FONDECYT, grant 1262503. GI is supported by a Juan de La Cierva fellowship (JDC2024-053550-I). CK acknowledges the grant EUR TESS N°ANR-18-EURE-0018 in the framework of the Programme des Investissements d'Avenir. DMR is supported by Tamkeen under the NYU Abu Dhabi Research Institute grant CASS. X.H. acknowledges the support by the National Natural Science Foundation of China under Grant No. 1237305.
\fast is a Chinese national mega-science facility, operated by National Astronomical Observatories of the Chinese Academy of Sciences (NAOC). 
Data used in this work can be obtained by sending a request to the FAST data centre. The Neil Gehrels Swift Observatory is a NASA/UK/ASI mission.
\end{acknowledgements}

\bibliographystyle{aa}
\bibliography{biblio}

\end{document}